\documentclass[11pt,twocolumn,prb,superscriptaddress,reprint]{revtex4-1}

\usepackage{graphicx,amsmath,amssymb,color}
\usepackage{tabularx, ctable}

\usepackage{ulem}
\usepackage{bm}
\usepackage{siunitx}
\usepackage{gensymb}

\usepackage[breaklinks=true,colorlinks,allcolors=blue]{hyperref}

\DeclareUnicodeCharacter{0301}{\'{e}}
\DeclareUnicodeCharacter{0301}{\'{0}}
\newcommand{\Supl}[1]{\textcolor{blue}{#1}}

\newcommand{\be}{\begin{eqnarray}}
\newcommand{\ee}{\end{eqnarray}}

\begin{document}

\title{High-mobility compensated semimetals, orbital magnetization and umklapp scattering in bilayer graphene moir\'e superlattices}

\author{A.L. Shilov}
\affiliation{Department of Materials Science and Engineering, National University of Singapore, 117575 Singapore}
\affiliation{Institute for Functional Intelligent
Materials, National University of Singapore, Singapore, 117575, Singapore}

\author{M.A. Kashchenko}
\affiliation{Programmable Functional Materials Lab, Center for Neurophysics and Neuromorphic Technologies, 127495}

\author{Pierre Pantale\'on}
\affiliation{Imdea Nanoscience, Faraday 9, 28015 Madrid, Spain}

\author{M. Kravtsov}
\affiliation{Department of Materials Science and Engineering, National University of Singapore, 117575 Singapore}
\affiliation{Institute for Functional Intelligent
Materials, National University of Singapore, Singapore, 117575, Singapore}

\author{A. Kudriashov}
\affiliation{Department of Materials Science and Engineering, National University of Singapore, 117575 Singapore}
\affiliation{Institute for Functional Intelligent
Materials, National University of Singapore, Singapore, 117575, Singapore}

\author{Z. Zhan}
\affiliation{Imdea Nanoscience, Faraday 9, 28015 Madrid, Spain}

\author{T. Taniguchi}
\affiliation{International Center for Materials Nanoarchitectonics, National Institute of Material Science, Tsukuba 305-0044, Japan}
\author{K. Watanabe}
\affiliation{Research Center for Functional Materials, National Institute of Material Science, Tsukuba 305-0044, Japan}

\author{S. Slizovskiy}
\affiliation{School of Physics and Astronomy, University of Manchester, Manchester, UK}

\author{K. S. Novoselov}
\affiliation{Institute for Functional Intelligent Materials, National University of Singapore, Singapore, 117575, Singapore}

\author{V.I. Fal’ko}
\affiliation{School of Physics and Astronomy, University of Manchester, Manchester, UK}

\author{F. Guinea}
\affiliation{Imdea Nanoscience, Faraday 9, 28015 Madrid, Spain}
\affiliation{Donostia International Physics Center, Paseo Manuel de Lardiz ́abal 4, 20018 San Sebastian, Spain}

\author{D.A. Bandurin$^{*}$}
\affiliation{Department of Materials Science and Engineering, National University of Singapore, 117575 Singapore}


\begin{abstract}

Twist-controlled moir\'e  superlattices (MS) have emerged as a versatile platform in which to realize artificial systems with complex electronic spectra. Bernal-stacked bilayer graphene (BLG) and hexagonal boron nitride (hBN) form an interesting example of the MS that has recently featured a set of unexpected behaviors, such as unconventional ferroelectricity and electronic ratchet effect. Yet, the understanding of the BLG/hBN MS electronic properties has, at present, remained fairly limited. Here we develop a multi-messenger approach that combines standard magnetotransport techniques with low-energy sub-THz excitation to get insights into the properties of this MS. We show that BLG/hBN lattice alignment results in the emergence of compensated semimetals at some integer fillings of the moir\'e bands separated by van Hove singularities where Lifshitz transition occurs. A particularly pronounced semimetal develops when 8 electrons reside in the moir\'e unit cell, where coexisting high-mobility electron and hole systems feature a strong magnetoresistance reaching 2350$\%$ already at $B=0.25$~T. Next, by measuring the THz-driven Nernst effect in remote bands, we observe valley splitting, pointing to an orbital magnetization characterized by a strongly enhanced effective $g_\mathrm{v}$-factor of 340. Last, using THz photoresistance measurements, we show that the high-temperature conductivity of the BLG/hBN MS is limited by electron-electron umklapp processes. Our multi-facet analysis introduces THz-driven magnetotransport as a convenient tool to probe the band structure and interaction effects in vdW materials and provides a comprehension of the BLG/hBN MS. 

\begin{center}
$^{*}$ Correspondence to: dab@nus.edu.sg
\end{center}
\end{abstract}

\maketitle
When atomically flat low-dimensional materials are brought into van der Waals (vdW) proximity with one another, they form a long-wavelength moir\'e superlattice if their crystal lattices are similar and properly aligned. The periodicity of the superlattice, together with interlayer hybridization, may lead to a significant alteration in the band structure of the parent materials, giving rise to a plethora of remarkable phenomena. Self-recurring energy spectra~\cite{Ashoori_superlattice,Ponomarenko2013, Dean-BLG-moire}, topological valley transport~\cite{Gorbachev2014}, high-temperature Brown-Zak oscillations~\cite{Rosh_BZ}, unconventional superconductivity~\cite{Cao2018}, correlated insulator states~\cite{Cao2018b}, integer and fractional Chern insulators~\cite{Spanton_Chern,Efetov_Chern, Yonlong_Chern}, orbital ferromagnetism~\cite{DGG_mag, Efetov_mag}, and quantized anomalous Hall effects~\cite{Serlin_mag} are just a few examples of the important phenomena discovered in moir\'e superlattices within recent years~\cite{Allan_Andrei_TBGreview, ACS_Nano_moire_review}.
\begin{figure*}[ht!]
  \centering\includegraphics[width=\linewidth]{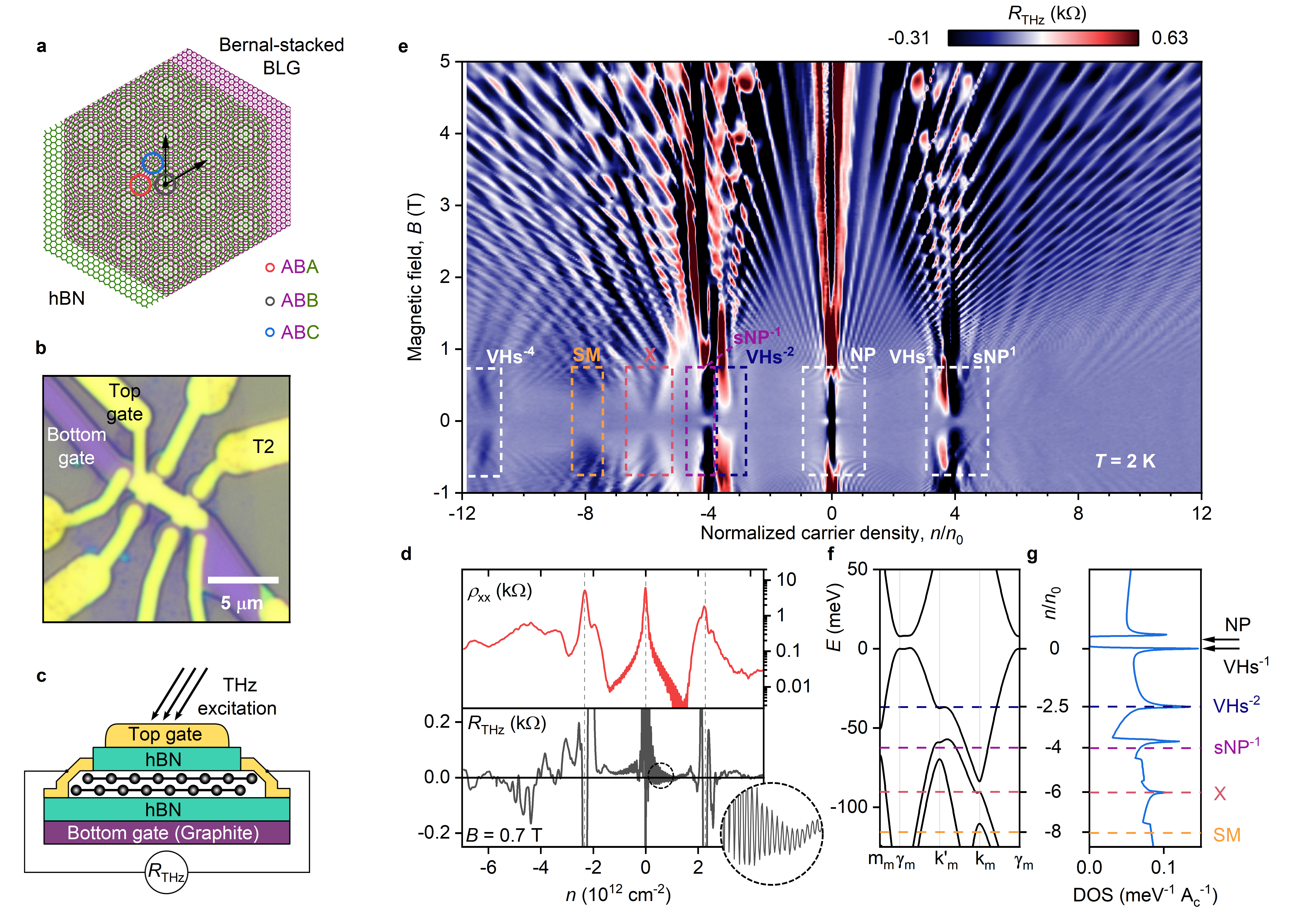}
    \caption{\textbf{BLG/hBN moir\'e superlattice.} \textbf{a,} Schematic of the moir\'e pattern formed between BLG and hBN. \textbf{b,} Optic image of the sample. The terahertz radiation is coupled to the structure via the top gate and the source (S) contacts.  \textbf{c,} Schematic of the sample: BLG is encapsulated between two slabs of hBN only one of which was aligned with the BLG lattice. $R_\mathrm{THz}$ is measured in a four-terminal configuration using a low-noise double-modulation technique (\Supl{Supplementary Information}).  \textbf{d,} Comparison of $\rho_{xx}(n)$ and $R_\mathrm{THz}$ at $B=0.7$ T. The inset reveals prominent oscillating features in $R_{\text{THz}}(n)$ that are less visible in $\rho_{xx}$ due to a strong background. Gray dashed lines indicate the positions of the main and secondary NPs.  \textbf{e,} $R_{\text{THz}}$ as a function of magnetic field and carrier density, measured at $T = 2$ K. The notation $4n_0 = 2.33\cdot10^{-12}$cm$^{-2}$ refers to the single miniband's filling.  \textbf{f-g,} Theoretical band structure and the DOS of the BLG/hBN superlattice calculated for the K high symmetry point of the original BLG BZ. Blue, purple, red, and orange dashed lines mark the hole-side 2$^{\text{nd}}$ VHs, 1$^{\text{st}}$ sNP, 3$^{\text{rd}}$ VHs, and the semimetallic state respectively.}
    \label{fig:F1}
\end{figure*}

Graphene-based superlattices represent a particularly important class, serving as a versatile platform for the design and exploration of novel electronic systems characterized by unique physical properties~\cite{Volodya_moire_review,Allan_Andrei_TBGreview, ACS_Nano_moire_review}. The most extensively studied case within this category is the moir\'e structure that forms when monolayer graphene (MLG) is deposited atop hBN, an atomically flat insulator with a hexagonal lattice nearly identical to that of graphene yet with a slightly larger crystal constant~\cite{Ashoori_superlattice,Ponomarenko2013}. In contrast, its not less interesting counterpart based on BLG/hBN structure has received considerably less attention, even though it was this structure that initially unveiled the existence of magnetic minibands in the energy spectrum of moir\'e superlattices~\cite{Dean-BLG-moire}. Recently, this system facilitated the observation of fractional quantization within these minibands~\cite{Spanton_Chern} and featured spontaneous polarization, giving rise to unconventional ferroelectricity~\cite{Qiong_ferro1,Ferro_BLG} and the electronic ratchet effect~\cite{Qiong_ratchet}. Nonetheless, despite these few observations and theoretical analysis~\cite{Volodya_BLG-hbn,Koshino_BLG-hbn,Pierre-BLG-hbn}, a comprehensive understanding of BLG/hBN superlattice's electronic properties, especially of its remote energy bands, has been missing. The scarcity of experimental data highlights the challenges associated with fabricating perfectly aligned, low-disorder BLG/hBN moir\'e structures~\cite{Top_currents_BLG-hbn}, which are necessary to address this inquiry. 

In this work, we produced such a high-quality perfectly aligned heterostructure and applied a multi-messenger approach that combines standard magnetotransport techniques with low-energy sub-THz excitation to gain deeper insights into the properties of this MS. We found that BLG/hBN alignment results in a series of compensated semimetals at some integer fillings of the moir\'e bands separated by van Hove singularities where Lifshitz transition converts electron Fermi pockets to those of hole type. A particularly pronounced semimetal emerges when 8 electrons reside in the moir\'e unit cell, where large-area coexisting electron and hole Fermi surfaces feature a giant magnetoresistance reaching 2350 $\%$ at low magnetic field $B=0.25$~T owing to intrinsically high carrier mobility. Next, by measuring the THz-driven Nernst effect in the remote bands, we observe giant valley splitting at $n/n_0=-6$, pointing to a strong orbital magnetization with an enhanced effective $g_\mathrm{v}$-factor of 340. Last, using THz photoresistance measurements, we show that electron-electron umklapp processes are the dominant source of resistance in this system. 

\textbf{Device and multi-messenger THz-driven magnetotransport.} Our sample was fabricated from Bernal-stacked BLG encapsulated by hBN slabs using a standard dry-transfer technique described elsewhere. As encapsulants, we used a large-area hBN flake that was unintentionally cracked into two pieces during exfoliation. One piece was aligned with respect to the BLG straight edge using an optical microscope equipped with micromanipulators and a high-precision rotation stage; this piece served as a top-gate dielectric. The second piece was intentionally misaligned by 30$\degree$ with respect to the BLG and served as a bottom-gate dielectric. The use of a cracked hBN flake ensured that only one of the pieces forms a moir\'e superlattice with the BLG flake eliminating the possibility of a double-moir\'e structure formation~\cite{Double-moire}. We note, in passing, that similar structures were reported to exhibit unconventional ferroelectricity~\cite{Qiong_ferro1} - the latter was notably absent in our devices. The obtained heterostructure was released on a few-layer graphite strip serving as a back gate. We then used standard electron-beam lithography, selective reactive ion etching, and thin film metal deposition to pattern top and bottom gates and contact leads (Fig. \ref{fig:F1}b,c). The top gate and one of the contact leads were connected to a broadband antenna enabling incident THz radiation to the funnel into the sample~\cite{Bandurin_dual,Bandurin2018,Gayduchenko2021}. The sample was mounted in the chamber of the magneto-optical variable temperature (1.8-300~K, 7~T) cryostat and exposed to incident sub-terahertz radiation ($f_{\text{THz}}=0.14~$THz) through the system of lenses and mirrors. All the measurements shown in the main text were performed at a zero displacement through simultaneous control of both top and bottom gate voltages, $V_\mathrm{tg}$ and $V_\mathrm{tg}$.

To explore the electronic properties of our device, we developed a multi-messenger technique that allowed us to simultaneously record longitudinal and transverse resistances of our sample: $\rho_\mathrm{xx}$ and $\rho_\mathrm{xy}$, as well as to measure the THz-driven photoresistance $R_\mathrm{THz}=R_\mathrm{on}-R_\mathrm{off}$ (here $R_\mathrm{off}$ and $R_\mathrm{on}$ are the resistances of the sample in the dark and upon THz exposure), and the photovoltage, $V_\mathrm{ph}$, built-up across the sample in response to incident radiation. While the photoresistance is a convenient probe for exploring quantization and scattering effects, as it isolates and highlights the contributions that are susceptible to changes in the electronic temperature, THz-induced $V_\mathrm{ph}$, mainly governed by the thermoelectric response~\cite{Cai2014,Bandurin_dual,Koppens_THz} (see below), is highly sensitive to the changes of the Fermi surface topology, and thus can be used to probe effects occurring close to the Lifshitz transitions~\cite{abrikosov2017fundamentals}. 

To demonstrate the advantages of our THz-driven approach, in Fig.~\ref{fig:F1}d we compare the $\rho_\mathrm{xx}$ and $R_\mathrm{THz}$ dependencies on the gate-induced carrier density $n$ measured at a relatively small magnetic field $B=0.7~$T, applied perpendicular to the sample plane. While $\rho_\mathrm{xx}$ captures only dominant transport features - main and secondary neutrality points (NP) as well as the quantum oscillations in the main moir\'e band, $R_\mathrm{THz}$ resolves the oscillatory pattern across the whole doping range. The high resolution of this approach is encoded in the suppression of the quantum oscillations under a THz-induced increase of the electronic temperature which, at small radiation power, has a weak effect on the smooth non-oscillating resistivity background. 
\begin{figure*}[ht!]
  \centering\includegraphics[width=\linewidth]{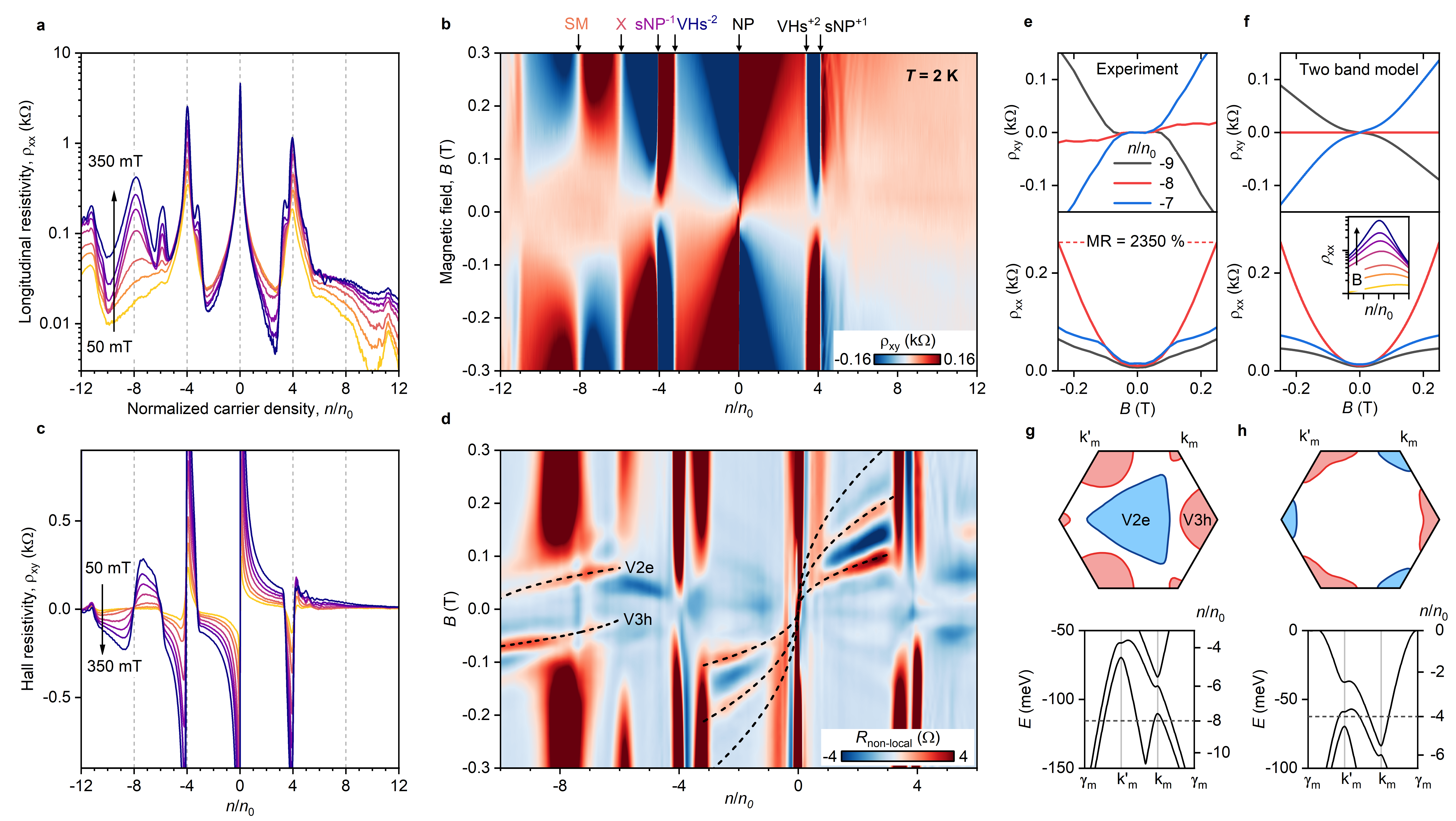}
    \caption{\textbf{Low-field magnetotransport and compensated semimetals.} \textbf{a,} Longitudinal resistance $\rho_\mathrm{xx}$ for $B$-fields ranging from 50 mT (orange) to 350 mT (blue). \textbf{b,} Hall resistance $\rho_\mathrm{xy}$ plotted against magnetic field and carrier densities \textbf{c,} Profiles of Hall $\rho_\mathrm{xy}$ vs $n/n_\mathrm{0}$ for $B$-fields ranging from 50 mT (orange) to 350 mT (blue). \textbf{d,} Non-local resistance map in a TMF configuration. Traces within the range of $-4 < n/n_0 < 4$ correspond to semi-classical orbits of carriers in the first pair of minibands. At higher negative fillings, traces marked as V2e and V3h reveal coexisting electron-like and hole-like Fermi surfaces. Results, presented in (a-d), were obtained at $T = 2$~K. \textbf{e-f,} Experimentally recorded $\rho_\mathrm{xy}(B)$ and $\rho_\mathrm{xx}(B)$ curves along with corresponding two-band conduction simulations near the compensation point. Carriers mobility of $\mu \sim 20\frac{\text{m}^2}{\text{V}\cdot\text{s}}$ was obtained from fitting the experimental data. \textbf{g-h,} Hole-like (red) and electron-like (blue) Fermi-pockets in the mini-BZ of the compensated semimetallic states at $n/n_0 = -8$ and $n/n_0 = -4$, respectively.}
    \label{fig:F2}
\end{figure*}

\textbf{Landau fan of the BLG/hBN superlattice.} Figure~\ref{fig:F1}d maps the $R_\mathrm{THz}$ dependence on $B$ and $n$ normalized to the density needed to fill the superlattice miniband $4n_\mathrm{0}=2.33\times10^{12}~$cm$^{-2}$ that we determine through the position of the sNP (Fig.~\ref{fig:F1}d) which, in turn, translates to the superlattice period of $\lambda = 14.08$ nm and alignment angle of $\theta\approx 0\degree$. The Landau fan in Fig.~\ref{fig:F1}e observed through $R_\mathrm{THz}$ adopts oscillating behavior resembling Hofstader's butterfly pattern and revealing Landau quantization of the first miniband at filling factors $\nu = 4m, m =1, 2, 3, ...$, characteristic of the BLG bands. Here positive $R_\mathrm{THz}$ is associated with the minima in $\rho_\mathrm{xx}$ and thus can serve as a guide to track the full filling of the Landau levels (LLs). Close to both secondary NPs labeled as sNP$^{-1}$ and sNP$^{1}$, the fan reveals the emergence of pronounced horizontal Hofstadter patterns at rational $\phi / \phi_0$ where $\phi$ is the magnetic flux through the superlattice unit cell and $\phi_0=h/e$ is the flux quantum (where $e$ is the elementary charge and $h$ is the Planck constant).  At $n/n_\mathrm{0}>4$ and $B>1.5~$T, $R_\mathrm{THz}$ reveals a continuation of the Landau quantization from the main band indicating the recovery of the unperturbed BLG electronic spectrum. At smaller $B$, we observe unusual low-frequency magnetooscillations with extrema that cannot be extrapolated to a single band filling at $B=0~$T. The origin of these oscillations is unknown but we tentatively attribute them to inter-subband scattering oscillations recently uncovered in twisted bilayer and double bilayer graphene~\cite{Isabelle_MISO, Volodya_MISO}. An alternative scenario involves cyclotron gap closure due to inter-subband LL crossing in a multiband magnetotransport regime~\cite{LL_gap_closure}. At $n/n_\mathrm{0}<-4$, the $R_\mathrm{THz}(n,B)$ pattern is much richer: in addition to the original structure of the BLG quantized spectrum with four-fold degeneracy, a mixed oscillation pattern is observed across the whole range $-12<n/n_\mathrm{0}<4$. Fast Fourier Transform (FFT) analysis reveals the presence of multiple fundamental frequencies with varying dependence on $n/n_\mathrm{0}$ making it difficult to disentangle the structure of the remote bands. 

\textbf{Low-field magnetotransport and high-mobility compensated semimetals.} To further the understanding of the complex energy spectrum of BLG/hBN superlattice, we now focus on the magnetic field range below $0.5~$T. In this range, the sign of $R_\mathrm{THz}$ can serve as an indicator of the NPs and van Hove singularities (vHS). Indeed, as the thermal broadening of the electron distribution causes a drop in resistivity at the neutrality points, negative $R_{\text{THz}}$ is expected to appear close to $n/n_0=0,\pm4$ in agreement with our data. On the contrary, we observe that NPs are surrounded by a region with strong positive $R_\mathrm{THz}$ at $n/n_\mathrm{0}=\pm0.2$ and $\pm3.2$. We attribute these features, not apparent in standard resistivity measurements, to van Hove singularities where THz-driven thermal broadening is overshadowed by the enhanced electron-hole scattering (see below). This leads to a positive $R_{\text{THz}}$, which enables us to distinguish vHS from NPs, where negative $R_{\text{THz}}$ is observed. Our interpretation is corroborated by the Hall effect measurements, which shows $\rho_\mathrm{xy}$ sign reversal at $n/n_0=\pm3.2$ (see below). Nevertheless, we point out that $R_{\text{THz}}$ provides somewhat more detailed diagnostics as it also reveals vHS (labeled as vHS$^{\pm1}$ close to the main gap edge at $n/n_\mathrm{0}=\pm0.2$ that is clearly obscured by large Hall coefficient in the $\rho_\mathrm{xy}(B)$ traces. Moreover, similar yet less pronounced positive peaks corresponding to the third vHS$^{-3}$ can be recognized at $n/n_0 = -6$ forming an "X"-shape feature marked as X in Fig.~\ref{fig:F1}d (see below). The described sequence of van Hove singularities is in perfect agreement with the theoretically calculated band structure of $0\degree$-aligned BLG/hBN superlattice represented in Fig. \ref{fig:F1}f,g, where the 2$^{\text{nd}}$ and the 3$^{\text{rd}}$ VHs are denoted by blue and red lines, respectively. Remarkably, at these points, not only does the density of states (DOS) function experience a sharp surge, but also the type of the carriers changes abruptly.

While the $R_{\text{THz}}$ captures well the emergence of vHS and NPs in BLG/hBN superlattice, it is instructive to compare it to conventional low-field magnetotransport.  Figure~\ref{fig:F2}a shows $\rho_{xx}$ dependencies on $n/n_0$ and reveals a series of maxima at the main and secondary NPs, along with small satellite peaks that appear at the secondary vHs at finite ($\sim$ 300 mT) magnetic field. More intriguingly, a prominent peak in $\rho_{xx}$ emerges at $n/n_0 = -8$ when a small ($\sim 200$ mT) magnetic field is applied. While this feature can be naively interpreted as the occurrence of the tertiary Dirac point~\cite{Tertiary_DP}, our experimental results challenge this intuition, as the photoresistance we recorded close to $n/n_0 = -8$ was remarkably small compared to the strong negative $R_{\text{THz}}$ at the main NP.

To better understand the magnetotransport properties of our device, we measured the Hall resistance $\rho_\mathrm{xy}$ at various carrier densities and magnetic fields  (Fig.~\ref{fig:F2}b). In this plot, $\rho_\mathrm{xy}$ changes its sign at specific points along the carrier density axis, corresponding to previously identified NPs and VHs. The behaviour of $\rho_\mathrm{xy}$ beyond the first pair of minibands is remarkably distinct at positive and negative fillings. At $n>4n_0$ the signal is small and lacks any visible singularities. However, at $n<-4n_0$ there exists a wide range of carrier densities near $n/n_0 =-8$ where the Hall resistance is extremely small at zero magnetic field and increases by orders of magnitude when subjected to finite magnetic field ($B>200$ mT). The rapid growth of $\rho_\mathrm{xy}$ in this region with magnetic field changing from 50 mT to 350 mT is more prominently visible in the $\rho_\mathrm{xy}$ profiles in fig. 2c. The unusual shape of $\rho_\mathrm{xy}$ is another distinguishing aspect of this feature: the sign change of $\rho_\mathrm{xy}$ at $n/n_0 = -8$ is characterized by a relatively smooth transition with a finite slope, in contrast to the abrupt change observed at the main NP.
\begin{figure*}[ht!]
  \centering\includegraphics[width=0.8\linewidth]{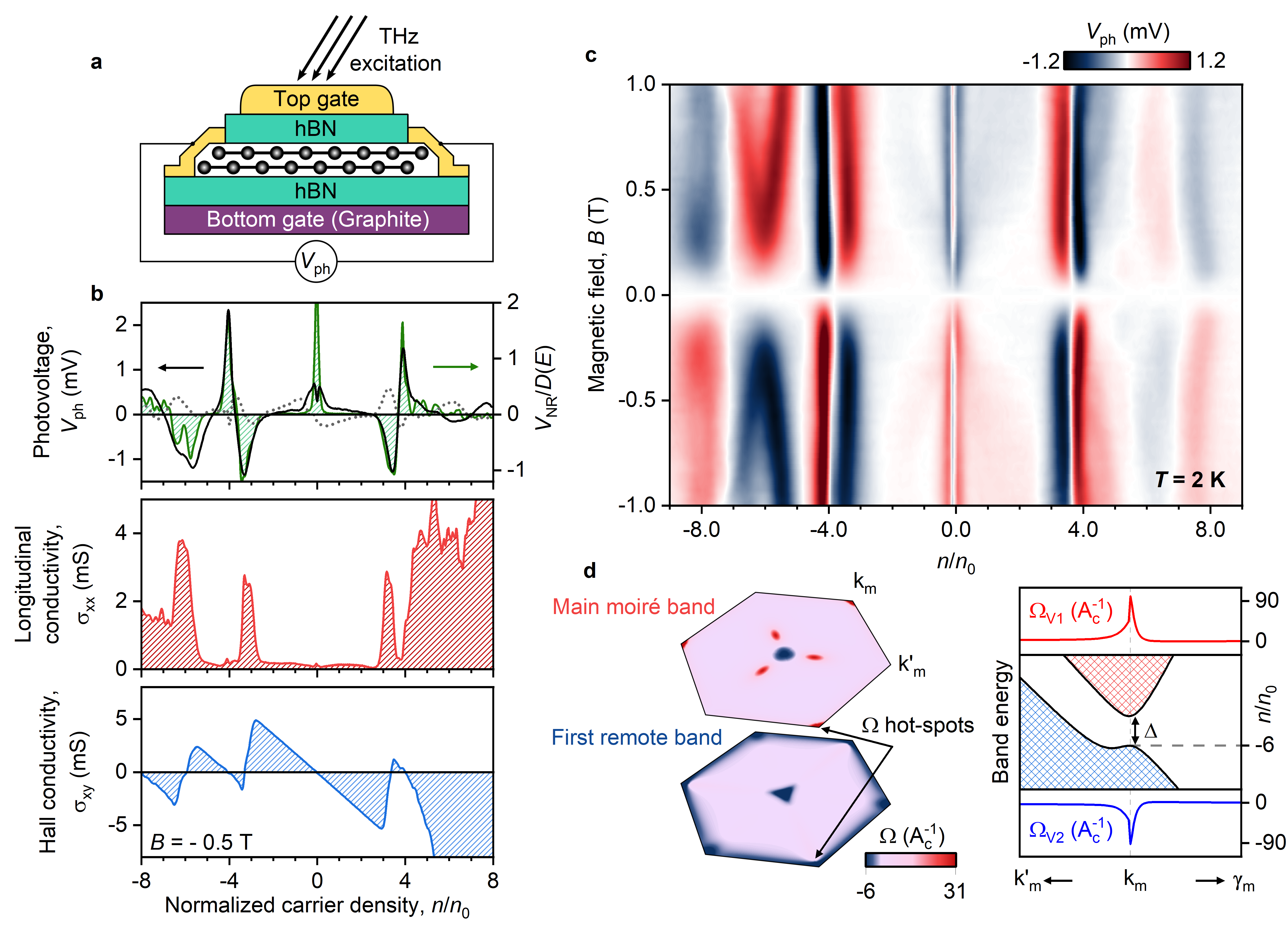}
    \caption{\textbf{THz-driven Nernst effect and large orbital magnetic moment.} \textbf{a,} Schematic of the photovoltage measurement configuration. $V_\mathrm{ph}$ is measured between the grounded source channel and the open-circuit drain channel located on the opposite side of the sample~\cite{Bandurin_dual}. \textbf{b,} Symmetric (black) and antisymmetric (grey) with respect to the $B-$field components of $V_\mathrm{ph}$ plotted atop longitudinal conductivity $\sigma_\mathrm{xx}$ (red), and Hall conductivity $\sigma_\mathrm{xy}$ (blue). $B = 0.5$~T. The green line shows the expected Nernst photovoltage normalized to the DOS  and calculated using \eqref{V_NR}. \textbf{c,} Symmetric part of $V_\mathrm{ph}$ mapped against the $B-$field and $n$. A splitting of the vHs is observed at $n/n_0 =-6$, corresponding to the orbital magnetic moment of $\sim 170\mu_B$. \textbf{d,} Theoretically calculated Berry curvature distribution at the K-valley of the original BLG's BZ in the main and first remote moir\'e bands. Berry curvature hot spots appear in the regions of close proximity of the bands at $k_\mathrm{m}$-points of the mini-BZ.} 
    \label{fig:F3}
\end{figure*}

To unravel the origin of the observed peculiarities, we applied the transverse magnetic focusing (TMF) technique that has become an important tool to probe the band structure of clean electronic systems including those of MS~\cite{tsoi1974focusing,TMF_Pablo,TMF_Volodya}  (\Supl{Supplementary Information}). TMF measurements reveal the coexisting electron-like and hole-like resonances in the vicinity of $n/n_0 = -8$ (corresponding traces are labeled as V2e and V3h respectively). The simultaneous appearance of multiple Fermi surfaces implies the overlap of the minibands and the semimetallic nature of the observed magnetoresistance (MR). To support this concept, we probed the resistance dependence on the magnetic field in the vicinity of $n/n_0 = -8$. The $\rho_{xy}(B),\;\rho_{xx}(B)$ curves exhibit behavior typical for semimetals, with MR reaching 2350 $\%$ at $n/n_0 = -8$, marking it as a compensation point of a semimetallic state. This conclusion is corroborated by the two-band conduction model that captures both the transition in the $\rho_{xy}$ slope and the enhancement of MR while crossing the compensation point (Fig.~\ref{fig:F2}e). We note in passing, that such a large $MR=(\mu B)^2$ corresponds to the high carrier mobility of $\mu=2\times10^5$ cm$^2$/Vs. Furthermore, this description aligns with the theoretically calculated band structure that predicts the coexistence of large electron and hole pockets at $n/n_0 = -8$ (Fig.~\ref{fig:F2}f).

Another valuable insight offered by the band structure is that, although the secondary NPs closely resemble conventional Dirac points, they, in fact, manifest as additional compensated semimetallic states with low carrier densities (Fig.~\ref{fig:F1}g). Our findings corroborate this prediction, as we recorded an unexpectedly low $R_{\text{THz}}$ at those points at $B = 0$. On the contrary, we observed a large negative $R_{\text{THz}}$ at the main NP even at low magnetic fields.


\textbf{THz-driven Nernst effect and orbital magnetization in remote bands.} Thermoelectric response can provide additional insights into the electronic properties of MS as they contain information on the interaction effects or phenomena that emerge close to Lifhitz transitions~\cite{Gosh_thermo,Ivan_thermo,Moriya2020,Paul2022}. The introduced multi-messenger technique allowed us to study the thermoelectric response of the BLG/hBN MS simultaneously with magnetotransport and THz-driven photoresistance. To this end, we leveraged on the fact that in our antenna-coupled devices, the radiation is coupled between the gate and source terminals, and thus there is a strong asymmetry in both the THz-induced high-frequency currents and carrier density profiles that leads to spatial asymmetric distribution of the electronic temperature~\cite{Cai2014,Bandurin_dual,Bandurin2018,Gayduchenko2021,Koppens_THz}. This, in turn, results in the dominance of thermoelectric effects over other rectification mechanisms~\cite{Bandurin_dual}. When the magnetic field is applied, the thermoelectric photovoltage is decomposed into the THz-driven Seebeck and Nernst components ($V_{SB}$ and $V_{NR}$ respectively) which differ by their reaction to the magnetic field. Using the Mott relation, one can obtain
\begin{equation}\label{V_SB}
    V_{SB}/D(E) \propto \rho_\mathrm{xx}\frac{d\sigma_\mathrm{xx}}{dn}+\rho_\mathrm{xy}\frac{d\sigma_\mathrm{yx}}{dn},
\end{equation}
\begin{equation}\label{V_NR}
    V_{NR}/D(E) \propto \rho_\mathrm{xx}\frac{d\sigma_\mathrm{xy}}{dn}+\rho_\mathrm{xy}\frac{d\sigma_\mathrm{xx}}{dn},
\end{equation}
where $D(E)$ is the DOS, $\sigma_\mathrm{xx}$ and $\sigma_\mathrm{xy}$ are the longitudinal and transverse elements of the conductivity tensor. These expressions suggest that the Seebeck part of the photovoltage is symmetric with respect to the magnetic field, while the contribution from the THz-driven Nernst effect is antisymmetric. In Fig.~\ref{fig:F3}b we show that the photovoltage appears as a series of strong peaks of the opposite signs at NPs and vHS under a finite magnetic field ($B=0.5$ T). This behavior is consistent with eqs. \eqref{V_SB},\eqref{V_NR}, given the non-monotonic behavior of conductivity near those points. The green curve in Fig. 2b represents $V_{NR}/D(E)$ that was calculated using \eqref{V_NR} and plotted in arbitrary units. Comparing it to the asymmetric part of the measured photovoltage,  we conclude that the THz-driven Nernst effect plays the predominant role in our configuration and acts as an analog to the photo-Nernst effect~\cite{PhotoNernst,Hot_carrier_moire}. Figure~\ref{fig:F3}c shows the complete map of the asymmetric part of $V_\mathrm{ph}$ across various fillings and magnetic fields (\Supl{Supplementary Information}). Among multiple peaks and dips in this plot, the feature near the vHS at $n/n_0 = -6$ particularly stands out, as it exhibits a strong splitting upon increasing magnetic field. A remnant of this anomaly can be also recognized in the photoresistance map for the same range of fillings (see Fig.~\ref{fig:F1}e). 


To understand what is responsible for such a strong splitting of the $V_\mathrm{ph}$ peak (dip) at the vHS in the vicinity of $n/n_0 = -6$ we return to the band structure (Fig.~\ref{fig:F3}d) and notice a small gap between the main and the first remote bands at the $k_\mathrm{m}$-point of the mini-Brillouin zone (BZ). While the first remote is passing through a vHS at this point, the overlying main band edge resembles a gapped Dirac dispersion. The latter is intuitively expected to feature a strong Berry curvature $\Omega(\textbf{k})$ hot spot and the associated orbital magnetic moment $m(\textbf{k})$. However, the emergence of $\Omega(\textbf{k})$ and $m(\textbf{k})$ hot spots has been usually explored in a two-band model where symmetric conduction and valence bands are separated by a band gap. Can the band alignment shown in Fig.~\ref{fig:F3} lead to a similar $\Omega(\textbf{k})$ and $m(\textbf{k})$ divergence? To answer this question, we calculate $\Omega(\textbf{k})$ for those bands and plot them in Fig.~\ref{fig:F3}. A closer look at $k_\mathrm{m}$ points indeed reveals the development of $\Omega(\textbf{k})$ hot spots and suggests the emergence of respective $m(\textbf{k})$. As the latter is opposite for the two parent valleys of the BLG bands, it is thus natural to attribute the splitting in Fig.~\ref{fig:F3}c to the field-induced valley splitting with a largely enhanced $g_\mathrm{v}$-factor. By tracking the peak position with $B$, one could estimate $g_\mathrm{v}-$factor to be of the order of 340 through $\Delta E=2 m B = g_\mathrm{v} \mu_B B$  and the orbital magnetic moment of $m=g_\mathrm{v}\mu_B/2=170\mu_B$, where $\mu_B$ is the Bohr magneton and $\Delta E$ is taken from Fig.~\ref{fig:F1}g at corresponding fillings. To compare, we note that the found values are several times larger than those observed in the remote band of MLG/hBN superlattices~\cite{Moriya2020} and those at the main band edges of moire-free BLG~\cite{BLG_PC1,BLG_PC2} yet 3 times smaller than in the case of ABA trilayer graphene~\cite{Slizovskiy_ABA} measured using different techniques.

\begin{figure*}[ht!]
  \centering\includegraphics[width=0.95\linewidth]{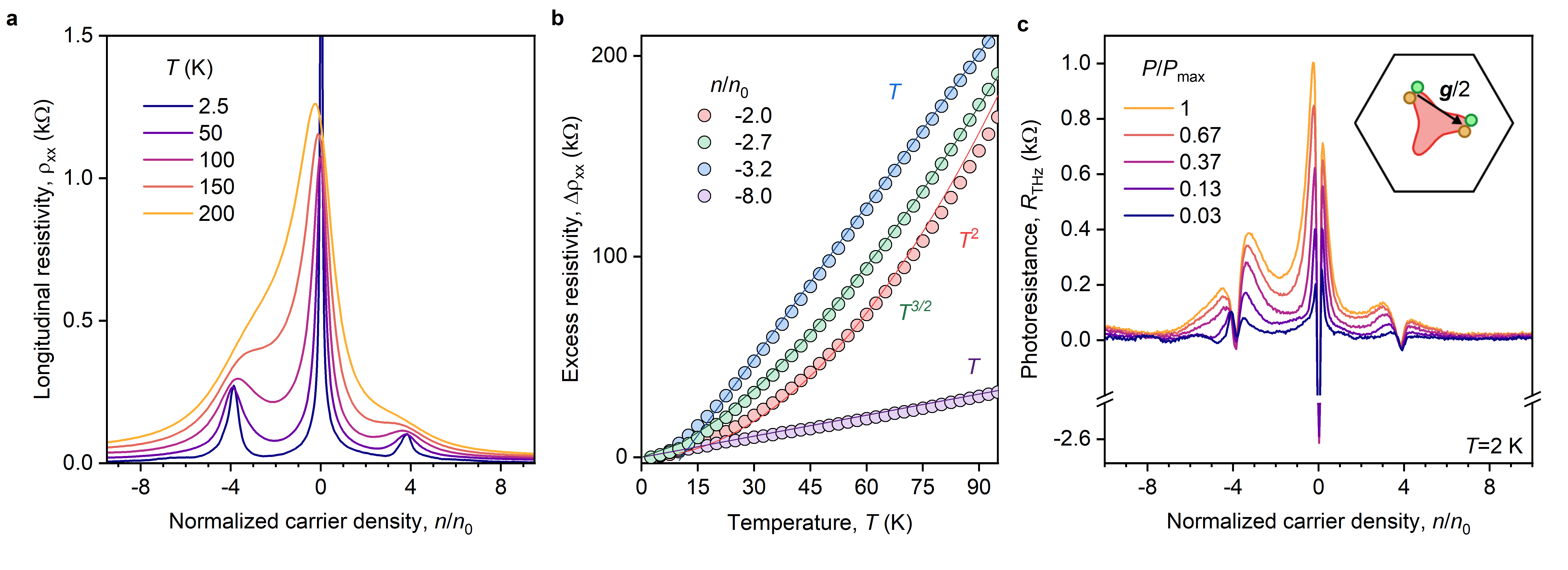}
    \caption{\textbf{Umklapp scattering in the BLG/hBN moir\'e bands.} \textbf{a,} Longitudinal resistivity $\rho_\mathrm{xx}$ vs $n/n_\mathrm{0}$, measured at $T$ varying from 2.5 K (blue) to 200 K (orange). \textbf{b,} Excess resistivity $\Delta \rho_\mathrm{xx}$ as a function of $T$ at various fillings, featuring a set of different functional behaviors $T^\alpha$ with $\alpha$ ranging from 1 to 2 depending on $n/n_\mathrm{0}$. Fitted curves are plotted in the same color as the corresponding raw data. \textbf{c,} $R_\mathrm{THz}$ as a function of $n/n_\mathrm{0}$, measured upon varying incident THz powers, $P$. The sample was kept at fixed $T=T_\mathrm{L}=2$~K. The inset illustrates the mechanism of umklapp scattering: after collisions, electrons appear on the opposite side of the Fermi surface so that the pair relaxes a reciprocal lattice unit vector $\textbf{\textit{g}}$.} 
    \label{fig:F4}
\end{figure*}

\textbf{High-\textit{T} conductivity and umklapp scattering.} Last, to complete the analysis of the electronic properties of the BLG/hBN MS properties, using our THz-driven technique we reveal the scattering mechanisms that are responsible for the resistivity of such system at elevated $T$. To this end, we first present the results of the conventional approach that relies on $\rho_\mathrm{xx}(T)$ measurements. Figure~\ref{fig:F4}a shows $\rho_\mathrm{xx}(n)$ dependencies recorded at various sample temperatures. The data reveals an increase in the sample resistivity as $T$ rises from 2.5~K to 200~K at all carrier densities except for the main NP.  We observed a rapid surge in $\rho_\mathrm{xx}$ at $-4 < n/n_0 < 4$, contrasting a more gradual increase of $\rho_\mathrm{xx}$ in the remote bands. Figure~\ref{fig:F2}b elaborates on this difference, showing the excess resistivity $\Delta\rho_\mathrm{xx}(T) = \rho_\mathrm{xx}(T) - \rho_\mathrm{xx}(2.5~\text{K})$ dependencies for various fillings. As we approach the midpoint of the first hole-side moire band, a clear $T^2$ growth with temperature is evident. However, a quadratic trend of $\Delta\rho_\mathrm{xx}(T)$ transitions to the linear behavior upon increasing $n/n_0$ passing through the filling at which $\Delta\rho_\mathrm{xx}\sim T^{3/2}$. At even higher negative fillings, excess resistivity significantly drops, remaining linearly dependent on temperature (green plot in Fig.~\ref{fig:F4}b). 

While the linear-in-$T$ contribution to $\rho_\mathrm{xx}(T)$ is commonly attributed to the electron-phonon scattering above the Bloch–Grüneisen temperature~\cite{Efetov_Phonons}, a $T^2$ scaling usually implies velocity-relaxing collisions among charge carriers~\cite{Wallbank2019}. The key difference in these contributions is that the former depends on the lattice temperature $T_\mathrm{L}$ whereas the latter on the electronic temperature, $T_\mathrm{e}$. In the conventional transport approach, the temperature of the sample sets both $T_\mathrm{L}$ and $T_\mathrm{e}$ making it nearly impossible to disentangle the dominant source of resistance from simple $\rho_\mathrm{xx}(T)$ fits especially when the $T^\alpha$ dependence with variable power $\alpha$ is observed. Our approach provides a convenient alternative to the transport analysis, as the sub-THz radiation, funneled inside the device can selectively increase $T_\mathrm{e}$ while keeping the lattice cold. Indeed, in graphene-based MS this is possible because at cryogenic $T_\mathrm{L}$ the electronic system, characterized by a vanishing specific heat capacity, can be efficiently decoupled from the lattice despite a relatively fast thermal relaxation.  In this context, photoresistance serves as a valuable measure, as it selectively filters out contributions sensitive only to electronic temperature $T_e$. Finite $R_{\mathrm{THz}}$ in the metallic regime (i.e., away from the NPs) implies the presence of the velocity-relaxing processes that appear due to inter-carrier collisions.

Figure~\ref{fig:F3} presents $R_{\mathrm{THz}}(n)$ curves at various incident 0.13~THz radiation powers $P$. In the metallic regime, we observe large positive $R_{\mathrm{THz}}(n)$ at $-4 < n < 4$ that grows with increasing $P$ (and, consequently, increasing $T_e$). Comparing it with the transport $\rho_\mathrm{xx}\sim T^2$ scaling, it is natural to assign this positive $R_{\mathrm{THz}}$ to the manifestation of umklapp scattering recently predicted to emerge in BLG MS~\cite{Volodya_BLG-umklapp}. Moreover, as $R_{\mathrm{THz}}$ is also strong and positive upon approaching the vHS where both $\alpha=1$ and 2 can be observed, we also conclude that in the main moir\'e band, the BLG/hBN MS is characterized by the interaction-limited conductivity. Last, the $R_\mathrm{THz}(n)$ plots reveal that the impact of electron-electron scattering mechanisms is substantially suppressed for $|n/n_0| > 8$, consistent with weak linear-in-$T$ $\Delta\rho_\mathrm{xx}(T)$ dependence measured at this filling pointing to the dominance of phonon scattering akin the case of moir\'e-free MLG which has vanishing THz photoresistance (\Supl{Supplementary Information}).

\textbf{Conclusions and outlook.} To sum up, we showed that a multi-messenger approach that combines standard magnetotransport techniques with low-energy sub-THz excitation can get deep insights into the properties of BLG/hBN MS. We demonstrated that BLG/hBN alignment results in the emergence of compensated semimetals at some integer fillings of the moir\'e bands separated by van Hove singularities where Lifshitz transition coverts electron Fermi pockets to those of hole type. A particularly pronounced semimetal develops at $n/n_0=-8$, where coexisting high-mobility electron and hole systems feature a giant magnetoresistance reaching 2350 $\%$ at a low magnetic field $B=0.25$~T. Next, by measuring the THz-driven Nernst effect in remote bands, we observed a strong valley splitting that we attribute to the presence of a topological magnetic moment with a strongly enhanced effective $g_\mathrm{v}$-factor of 340. Last, using THz photoresistance measurements, we show that the high-temperature conductivity of the BLG/hBN MS is limited by electron-electron umklapp processes in the main moire band which are suppressed in the remote bands. Our multi-facet analysis introduces THz-driven magnetotransport as an important tool to probe the electronic properties and interaction effects in vdW MS and provides a comprehensive understanding of the BLG/hBN superlattices. It would be interesting to expand such studies to small-angle twisted bilayer graphene devices to get further insight into their intriguing properties. 




\section*{Competing interests}
The authors declare no competing interests.



\clearpage
\bibliography{Bibliography.bib}

\newpage
\setcounter{figure}{0}
\renewcommand{\thesection}{}
\renewcommand{\thesubsection}{S\arabic{subsection}}
\renewcommand{\theequation} {S\arabic{equation}}
\renewcommand{\thefigure} {S\arabic{figure}}
\renewcommand{\thetable} {S\arabic{table}}
\end{document}